\documentclass[times,authoryear]{elsarticle}

\usepackage{jasr}
\usepackage{framed,multirow}

\usepackage{amssymb}
\usepackage{latexsym}

\usepackage[switch]{lineno}

\usepackage{url}
\usepackage{xcolor}
\definecolor{newcolor}{rgb}{.8,.349,.1}

\usepackage[citebordercolor=white]{hyperref}

\journal{Advances in Space Research}

\begin{document}

\verso{Yogesh \textit{et al.}}

\begin{frontmatter}

\title{Statistical insights on the decorrelation lengths of solar wind parameters at L1 point under varying conditions \tnoteref{correlation}}%

\author[1]{Yogesh \corref{cor1}}
\ead{yphy22@gmail.com}
\author[3,4]{Aparupa Apsara \snm{Baruah}}
\author[5]{Dibyendu \snm{Chakrabarty}}
\author[3,4]{Kalyan \snm{Bhuyan}}
\author[5,6]{Aakash \snm{Gupta}}
\author[1]{Gregory G. \snm{Howes}}
\affiliation[1]{organization=Department of Physics and Astronomy, University of Iowa,
                city=Iowa City, IA,
                postcode=52242,
                country={USA}}

\affiliation[3]{organization=Dept. of Physics, Dibrugarh University,
                city=Assam,
                postcode=786001,
                country=India}    
                
\affiliation[4]{organization=Centre for Atmospheric Studies,  Dibrugarh University,
                city=Assam,
                postcode=786001,
                country=India}

\affiliation[5]{organization=Physical Research Laboratory,
                city=Ahmedabad, Gujarat ,
                postcode=380009,
                country=India}

\affiliation[6]{organization=Indian Institute of Technology (IIT), city=Gandhinagar,postcode=382055, country=India}                

\begin{abstract}
Understanding the spatial coherence of solar wind plasma and magnetic field properties is essential for interpreting multi-spacecraft observations and for characterizing the large-scale structure of heliospheric transients. In this study, we quantify the spatial correlation of six key solar wind parameters—interplanetary magnetic field components, bulk flow speed, proton number density, and the alpha-to-proton abundance ratio—using simultaneous measurements from the \textit{ACE} and \textit{Wind} spacecraft as a function of their instantaneous separation distance. The analysis is performed separately for intervals of background solar wind, Interplanetary Coronal Mass Ejections (ICMEs), and Stream Interaction Regions (SIRs). The decay of the Pearson correlation coefficient with distance is modeled using an exponential function to infer characteristic de-correlation length scales. We find that the bulk solar wind speed is the most spatially coherent parameter in all regimes, while plasma composition exhibits the weakest coherence. Magnetic field coherence shows strong dependence on solar wind structure: ICMEs display near-unity correlations and the largest magnetic coherence scales, consistent with organized, flux-rope-like configurations, whereas SIRs exhibit reduced coherence—particularly in the north--south magnetic field component—reflecting compressed and turbulent plasma. The background solar wind exhibits intermediate behavior, with large-scale coherence in bulk plasma properties but shorter coherence lengths in magnetic fluctuations. These results provide a quantitative framework for distinguishing solar wind structures based on their spatial coherence properties and have important implications for multi-point solar wind studies and space weather applications.
\end{abstract}

\begin{keyword}
\KWD Keyword1\sep Keyword2\sep Keyword3
\end{keyword}

\end{frontmatter}


\section{Introduction}\label{sec:intro}

The Sun, as the primary energy source of the heliosphere, continuously emits electromagnetic radiation and a stream of charged particles known as the solar wind into interplanetary space. Temporal and spatial variability in the lower solar atmosphere gives rise to fluctuations in solar wind properties, which are further modified by convective distortions associated with flow inhomogeneities and variations in propagation speed with heliographic latitude and longitude \citep{Roberts1999}. Observations of the solar wind at 1~AU over several decades have demonstrated that its local plasma and magnetic field characteristics are strongly influenced by the solar cycle \citep{Hapgood1991}, with this influence extending throughout the heliosphere and manifesting in a wide range of heliophysical and geophysical phenomena \citep{Richardson2008}.

The solar wind mediates the transport of plasma and magnetic fields throughout the heliosphere and plays a central role in solar--terrestrial interactions. Its properties evolve continuously with radial distance from the Sun \citep{Parker1958, Schwartz1983, Richardson1995, Bruno2003, Perrone2019, Yogesh2026}. The bulk velocity and embedded magnetic field exhibit pronounced turbulent fluctuations \citep{Goldstein1995}, while the flow also contains large-scale coherent structures such as coronal mass ejections (CMEs; \citealp{Wimmer2006}) and stream interaction regions (SIRs), which form through interactions between fast and slow solar wind streams \citep{Richardson2018}. These large-scale structures strongly influence the acceleration, transport, and spatial redistribution of energetic particles in the heliosphere. Recent observations from the Aditya Solar Wind Particle Experiment (ASPEX)  \citep{Goyal2018} on board Aditya-L1 \citep{Seetha2017, Tripathi2022, Parate2025} have shown that interactions between multiple interplanetary CMEs can generate extended downstream interaction regions associated with particle energization, shock propagation, and cross-field redistribution of ions in the heliosphere \citep{Parashar2026}. In addition, quiet-time suprathermal ion populations measured by ASPEX-STEPS \citep{Sebastian2026} have been found to exhibit nearly isotropic directional behavior over timescales of a few days, providing observational support for the isotropy assumption commonly employed in Parker transport theory for energetic particle propagation \citep{Parker1965, Fisk2008, Fisk2012, Fisk2014, Gupta2025}. Both turbulence and macroscopic structures strongly influence the interaction of the solar wind with planetary magnetospheres, driving phenomena such as auroral activity \citep{Legrand1985}, geomagnetic storms \citep{Cliver1996}, and ionospheric disturbances that can adversely affect communication and navigation systems. These effects highlight the importance of understanding solar wind variability in the broader context of space weather and near-Earth environmental conditions.

A fundamental approach to characterizing the impact of solar activity on the solar wind is to determine the spatial correlation or coherence length scale ($\lambda$) of in situ plasma and magnetic field fluctuations \citep{Wicks2009}. Two complementary techniques are commonly employed to estimate $\lambda$. The first applies the Taylor frozen-in-flow hypothesis \citep{Taylor1938} to single-spacecraft time series observations \citep[e.g.,][]{Smith2018, Ragot2022}, enabling the characterization of large-scale, slowly evolving structures typically observed over timescales of several hours \citep{Weimer2003, Wicks2010}. The second approach uses simultaneous measurements from multiple spacecraft to directly infer spatial correlation lengths from known spacecraft separations \citep[e.g.,][]{Matthaeus2005, Osman2007, Podesta2008}. This multi-spacecraft technique provides instantaneous spatial information without relying on assumptions regarding flow stationarity or advection speed, making it particularly valuable for studying evolving solar wind structures.

Understanding the spatial correlation of solar wind fluctuations is essential for interpreting the interplay between turbulence \citep[e.g.,][]{Goldstein1995, Pommois2001}, large-scale coherent structures, and the coronal processes that drive solar wind formation. Both turbulence and macroscopic solar wind structures exhibit solar cycle dependence \citep[e.g.,][]{Wicks2010}, and their spatial coherence directly influences the predictability of solar wind conditions at different heliospheric locations. Consequently, quantifying these spatial scales is a critical component of space weather modeling and forecasting efforts \citep{Temmer2021}.

The inner heliosphere is dominated by three primary large-scale plasma configurations whose distinct spatial structures govern space weather dynamics at 1 AU.  The ambient solar wind forms the continuous background matrix, containing structures that stretch radially across $\sim 0.01-0.05$ AU \citep{Viall2021}. Embedded within this background are ICMEs---transient eruptions consisting of a compressed, turbulent shock-sheath region followed by a coherent magnetic obstacle or flux rope---which exhibit an average radial width of roughly $ \sim 0.25-0.30$ AU at Earth's orbit \citep{Jian2006a}. In contrast, SIRs span a typical radial width of $\sim 0.20-0.30$ AU at 1 AU, widening as they propagate into the outer heliosphere to mature into co-rotating interaction regions (CIRs) \citep{Jian2006b}.  Although SIRs possess vast spatial extents, they exhibit significant small-scale variations in space and time. In this study, we utilize de-correlation lengths to characterize these structural variations across these different categories.

Previous studies of de-correlation lengths have primarily focused on the background solar wind, with limited attention to other solar wind categories such as ICMEs and SIRs. These two categories, however, are particularly important for space weather prediction. Moreover, compositional correlation lengths, such as those of helium abundance, have not been systematically explored in past work. In this study, we quantify the spatial correlation of several key solar wind parameters as a function of the instantaneous separation distance between spacecraft, expressed in Earth radii ($R_E$). The parameters analyzed include the interplanetary magnetic field components ($B_x$, $B_y$, $B_z$), the plasma bulk speed ($V$), the proton number density ($N_p$), and the helium abundance ratio ($A_{\mathrm{He}}$), defined as the ratio of alpha particle number density to proton number density ($n_{\mathrm{He}}/n_p$). Helium abundance is a fundamental compositional parameter of the solar wind and exhibits systematic variations under different solar wind conditions, including background solar wind \citep{Alterman2019, Yogesh2021, Ofman2024, Yogesh2024}, ICMEs \citep{Fu2018, Yogesh2022}, and SIRs \citep{Durovcova2019, Yogesh2023}. These variations reflect differences in solar source regions, coronal heating processes, and solar wind acceleration mechanisms. 

Other heavy-ion compositional parameters are not considered in this analysis because the \textit{Wind} spacecraft does not provide routine measurements of heavy ion species beyond helium, which limits direct compositional comparisons with the \textit{ACE} spacecraft. We employ simultaneous observations from \textit{ACE} and \textit{Wind} to compute instantaneous correlation coefficients for each parameter over a range of spacecraft separations extending from approximately 10 to more than 300~$R_E$ in Geocentric Solar Ecliptic (GSE) coordinates \citep{Hapgood1992}. The variation of correlation with separation distance is modeled using an exponential decay function of the form $a\,\exp(-x/b)$, where $a$ represents the extrapolated zero-separation correlation and $b$ denotes the e-folding correlation length. This approach provides quantitative estimates of the spatial coherence scales associated with turbulence and large-scale solar wind structures under different solar wind conditions.

The structure of the paper is as follows: Section \ref{sec:data} describes the data set used in this study, Section \ref{sec:results} presents the data analysis and key observational results, and Section \ref{sec:disc} provides the discussion, and section \ref{sec:conclusion} shows the conclusions drawn from the study.

\section{Data Section and event selection}\label{sec:data}

\subsection{Data}

In this study, interplanetary magnetic field (IMF) measurements are obtained from the Magnetic Field Investigation (MFI) instrument on board the \textit{Wind} spacecraft \citep{Lepping1995} and the Magnetic Field Experiment (MAG) on the \textit{ACE} spacecraft \citep{Smith1998}. Solar wind plasma parameters are derived from the Solar Wind Experiment (SWE) on \textit{Wind} \citep{Ogilvie1995} and the Solar Wind Electron Proton Alpha Monitor (SWEPAM) on \textit{ACE} \citep{McComas1998}. The native time resolutions of the plasma data are $\sim$92s for \textit{Wind}/SWE and $\sim$64s for \textit{ACE}/SWEPAM. 

While both the SWE on the \textit{Wind} spacecraft and the SWEPAM on the \textit{ACE} measure core solar wind plasma parameters, they rely on fundamentally different instrumental techniques. SWE utilizes Faraday Cup (FC) technology, which acts as a retarding potential analyzer by using modulated grid voltages to filter ions based on their energy-per-charge ($E/q$) and measuring the resulting bulk current at a collector plate. 
Conversely, SWEPAM employs curved-plate Electrostatic Analyzers (ESAs), which utilize a static deflection voltage to select and sweep through narrow $E/q$ and angular ranges, counting individual particle impacts via channel electron multipliers. Despite these operational differences, both instruments derive bulk velocity, density, and alpha particle abundance from the measured ion energy distribution functions. 

We analyze the three components of the magnetic field ($B_x$, $B_y$, and $B_z$), the solar wind bulk speed ($V$), the proton number density ($N_p$), and the helium abundance ratio ($A_{\mathrm{He}}$). All magnetic field components are expressed in the Geocentric Solar Ecliptic (GSE) coordinate system. To enable direct comparison and correlation between \textit{Wind} and \textit{ACE} observations, all datasets are resampled to a common cadence of 2~min using linear averaging. The instantaneous spacecraft separation is calculated as the magnitude of the three-dimensional position vector difference between \textit{Wind} and \textit{ACE}.

The SWE Faraday cups cannot reliably track the bulk ion population inside the magnetosphere \citep{Wilson2021}. Therefore, we use only Wind/SWE measurements flagged as good quality, which removes intervals contaminated by the Earth's magnetospheric plasma.

\subsection{Event Selection}

The dataset is categorized into three solar wind regimes: background solar wind, interplanetary coronal mass ejections (ICMEs), and stream interaction regions (SIRs). ICME intervals are identified using the \textit{Wind} ICME catalog\footnote{\url{https://wind.nasa.gov/ICME_catalog/ICME_catalog_viewer.php}}, which primarily lists magnetic obstacle intervals. A detailed description of the catalog and its identification criteria is provided by \citet{Chinchilla2019}. The inclusion or exclusion of the turbulent sheath region of ICMEs can significantly impact statistical results. Therefore, this study focuses exclusively on magnetic clouds, which are defined as magnetic obstacles (MOs) within the ICME catalog.

SIR events are selected from the catalog compiled by \citet{Chi2018}, which includes 866 SIRs observed between 1995 and 2016, along with their corresponding start and end times at the observing spacecraft. Further details regarding the identification methodology and event characteristics are described in \citet{Chi2018} and references therein. In this analysis, we include the entire SIR, capturing both the slow- and fast-wind boundaries. These boundaries are selected to fully encompass the compressional effects occurring on both sides of the interface. While the current study is limited to the SIR catalog spanning 1995 to 2016, this dataset will be extended in future work to incorporate newer catalogs \citep{Mayank2026}.

To isolate intervals of background solar wind, all time periods associated with ICMEs and SIRs are excluded from the analysis. In addition, a buffer interval of 12~hours before and after each ICME and SIR event is removed to minimize contamination from disturbed plasma conditions in the surrounding regions.

\section{Results}\label{sec:results}

The \textit{ACE} and \textit{Wind} spacecraft are two primary missions operating near the Sun--Earth L1 Lagrange point and are routinely used for monitoring upstream solar wind conditions relevant to space weather. The separation between the two spacecraft varies substantially over time, ranging from approximately 10~$R_E$ to more than 300~$R_E$, particularly during intervals when \textit{Wind} traverses Earth’s orbit. This variability in separation provides a unique opportunity to examine the spatial coherence of solar wind plasma and magnetic field structures over a wide range of spatial scales.

Understanding the degree of correlation between measurements obtained by \textit{ACE} and \textit{Wind} is essential for assessing the spatial and temporal coherence of solar wind structures and for evaluating the reliability of upstream observations used in space weather forecasting. Differences in spacecraft separation directly influence the extent to which measurements at one location can be considered representative of conditions elsewhere in the near-Earth solar wind.

In this section, we analyze the relationship between simultaneous observations from \textit{ACE} and \textit{Wind} using the full dataset. We quantify the dependence of the correlation between corresponding parameters on spacecraft separation distance and examine how this dependence varies across different solar wind regimes, including background solar wind, CMEs, and SIRs.

\subsection{Comparison Between ACE and Wind Observations}

\begin{figure}[h!]
\includegraphics[scale=0.7]{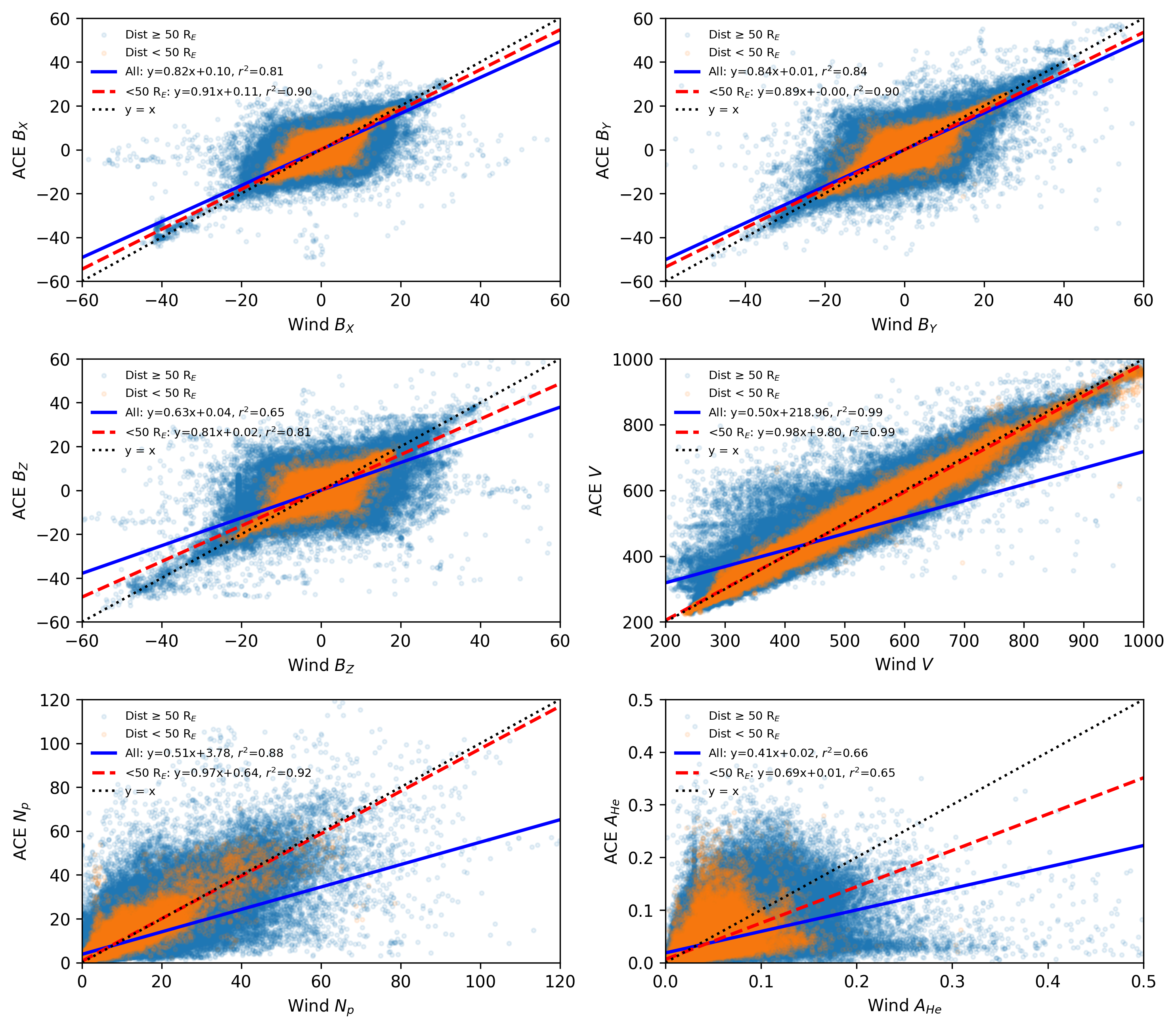}
\caption{Comparison of contemporaneous solar wind measurements from the ACE and \textit{Wind} spacecraft. The six panels display correlations for magnetic field components ($B_x, B_y, B_z$), flow velocity ($V$), proton density ($N_p$), and helium abundance ratio ($A_{He}$). The $y$-axis represents ACE observations, while the $x$-axis represents \textit{Wind} observations. Data points are separated into two categories based on spacecraft separation distance: orange dots indicate a separation of $<50\,R_E$, and light blue dots indicate a separation of $\ge 50\,R_E$. Linear least-squares regression lines are shown for the full dataset (solid blue line) and the close-proximity subset (dashed red line). The dotted black line represents the ideal unity slope ($y=x$). The corresponding linear fit equations and Pearson correlation coefficients ($r^2$) are displayed within each panel.}
\label{fig:ace_wind_corr}
\end{figure}

Figure~\ref{fig:ace_wind_corr} presents a comparative correlation analysis of contemporaneous magnetic field and plasma measurements obtained by the \textit{ACE} and \textit{Wind} spacecraft. Six key solar wind parameters are examined: the magnetic field components ($B_x$, $B_y$, $B_z$), the bulk solar wind speed ($V$), the proton number density ($N_p$), and the helium abundance ratio ($A_{\mathrm{He}}$). For each parameter, a linear least-squares regression of the form $y = mx + c$ is performed, and the strength of the relationship is quantified using the Pearson correlation coefficient ($r^2$). The analysis is conducted for the full dataset (denoted ``All'') and for a subset restricted to intervals when the spacecraft separation is less than $50~R_E$.

In Figure~\ref{fig:ace_wind_corr}, \textit{ACE} measurements are plotted along the $y$-axis and the corresponding \textit{Wind} measurements along the $x$-axis. Data points corresponding to separations smaller than $50~R_E$ are shown in orange, while those obtained at larger separations ($\geq 50~R_E$) are shown in light blue. Linear regression fits are indicated by red dashed lines for the close-separation subset and solid blue lines for the full dataset. The black dotted line represents the $y=x$ identity line, corresponding to perfect agreement between the two spacecraft. The scale of $50~R_E$ is chosen because flux ropes observed in the solar wind typically have widths of this order \citep{Borovsky2008}. Their study showed that the flux-rope size distribution spans a broad range, from approximately $10~R_E$ to $300$--$400~R_E$ with peak near $50~R_E$. Accordingly, in the following sections, we consider spatial scales within this range, consistent with the observations reported by \citet{Borovsky2008}.

A clear dependence of the correlation on spacecraft separation is evident across most parameters. For the magnetic field components, the correlations improve significantly when the spacecraft are within $50~R_E$. Both $B_x$ and $B_y$ exhibit strong correlations in the close-separation subset ($r^2 = 0.90$ for each), with regression slopes approaching unity ($m = 0.91$ and $m = 0.89$, respectively). The $B_z$ component, which is particularly relevant for space weather applications, also shows a marked improvement, with the correlation increasing from $r^2 = 0.65$ for the full dataset to $r^2 = 0.81$ for the close-separation subset, and the regression slope increasing from $m = 0.63$ to $m = 0.81$.

Among the plasma parameters, the solar wind bulk speed ($V$) exhibits the strongest overall correspondence between \textit{ACE} and \textit{Wind}. For the full dataset, the correlation is very high ($r^2 \approx 1$), indicating that both spacecraft capture the same large-scale velocity variations. However, the regression slope is substantially underestimated ($m = 0.50$), with a large offset ($c = 218.9$), suggesting that increased spacecraft separation introduces systematic differences in the measured velocity amplitudes. When the analysis is restricted to separations smaller than $50~R_E$, the agreement improves dramatically, with the correlation remaining high ($r^2 = 0.99$), the slope approaching unity ($m = 0.98$), and the offset decreasing to $c = 9.8$. This indicates that, at close separations, both spacecraft observe nearly identical velocity structures.

A similar behavior is observed for the proton number density ($N_p$). For the full dataset, the correlation remains relatively high ($r^2 = 0.88$), but the regression slope is modest ($m = 0.51$) and considerable scatter is present. Restricting the analysis to separations below $50~R_E$ results in a notable improvement, with the correlation increasing to $r^2 = 0.92$ and the slope approaching unity ($m = 0.97$). This suggests that density structures are less reliably sampled at larger inter-spacecraft separations.

In contrast, the helium abundance ratio ($A_{\mathrm{He}}$) exhibits the weakest correlation among the plasma parameters under both separation conditions. For the full dataset, the regression slope is low ($m = 0.41$) and the correlation is moderate. The close-separation subset shows improvement, with $r^2 = 0.65$ and $m = 0.69$, but the agreement remains poorer than for $V$ and $N_p$. This reduced correlation may reflect differences in instrumental sensitivity and measurement techniques between \textit{ACE} and \textit{Wind}, as well as the inherent variability of compositional parameters.

In addition, the parallelogram-like structure observed in the magnetic field components, as well as the irregular features in $A_{\mathrm{He}}$, may be related to whether the spacecraft separation is predominantly parallel or perpendicular to the local magnetic field. In addition, a few outliers corresponding to distinct structural signatures are also present, which may be associated with specific classes of events. A detailed investigation of these geometric effects and special outliers are beyond the scope of this study and will be addressed in future work.

Overall, the systematic variation of correlation coefficients and regression slopes with spacecraft separation indicates that the characteristic spatial scales of solar wind structures play a central role in producing differences between \textit{ACE} and \textit{Wind} observations. In the following subsection, we examine the dependence of correlation on separation distance in greater detail and investigate how this behavior varies across different solar wind regimes, including background solar wind, CMEs, and SIRs.

\subsection{Correlation Between Wind and ACE Observation as a Function of Spacecraft Separation in the Solar Wind }

\begin{figure}[h!]
\centering
\includegraphics[scale=0.6]{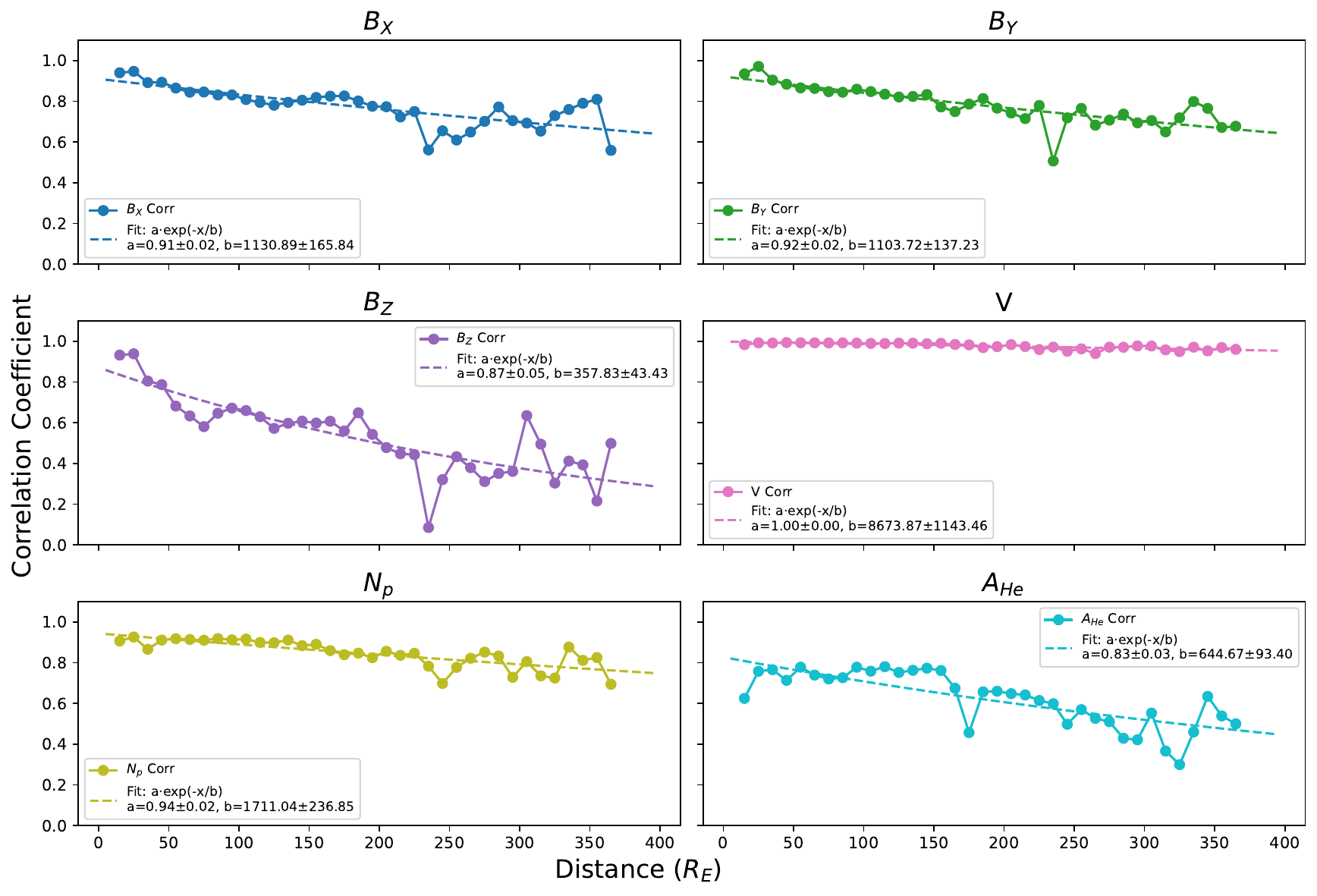}
\caption{Correlation evolution of plasma and magnetic field parameters with distance during backgroun solar wind intervals. Correlation coefficients for $B_x$, $B_y$, $B_z$, $V$, $N_P$, and $A_{He}$ are plotted against distance (in $R_E$). The exponential model $a \cdot \exp(-x/b)$ is fitted to the data, with best-fit parameters $a$ and $b$ displayed in each subplot.}
\label{fig:sw}
\end{figure}

We quantify the spatial correlation of six solar wind parameters measured by the \textit{ACE} and \textit{Wind} spacecraft as a function of their instantaneous separation distance, restricting the analysis to intervals of background solar wind. In Figure \ref{fig:sw}, for each 10 $R_E$ separation bin, the Pearson correlation coefficient ($r^2$) is calculated and its decay with distance (d) is modeled using an exponential function,
\begin{equation}
r(d) = a\,\exp(-d/b),
\end{equation}
where $a$ represents the extrapolated zero-separation correlation and $b$ denotes the characteristic de-correlation length scale.

The interplanetary magnetic field (IMF) components exhibit moderate to strong spatial correlations, with clear anisotropy among the field orientations. The radial ($B_x$) and transverse ($B_y$) components display comparable de-correlation lengths of $b = 1130.89 \pm 165.84~R_E$ and $b = 1103.72 \pm 137.23~R_E$, respectively. In contrast, the north--south component ($B_z$) decorrelates substantially more rapidly than the other IMF components, with a characteristic length of $b = 357.83 \pm 43.43~R_E$.

Among the plasma parameters, the bulk solar wind speed ($V$) is the most spatially coherent, characterized by an extrapolated correlation of $a = 1.00 \pm 0.00$ and an exceptionally large de-correlation length of $b = 8673.87 \pm 1143.46~R_E$. The proton number density ($N_p$) also remains highly correlated over large separations, yielding a de-correlation length of $b = 1711.04 \pm 236.85~R_E$. In comparison, the alpha-to-proton abundance ratio ($A_{\mathrm{He}}$) exhibits a shorter coherence scale of $b = 644.67 \pm 93.40~R_E$ and the lowest inferred zero-separation correlation, $a = 0.83 \pm 0.03$.

Overall, these results demonstrate that solar wind parameters span a broad range of spatial coherence scales. Bulk flow speed and proton density remain coherent over large distances, whereas IMF fluctuations—particularly in the $B_z$ component—are governed by more localized or rapidly evolving structures.

\subsection{Correlation Between Wind and ACE Observation as a Function of Spacecraft Separation in the ICMEs }

\begin{figure}[h!]
\centering
\includegraphics[scale=0.7]{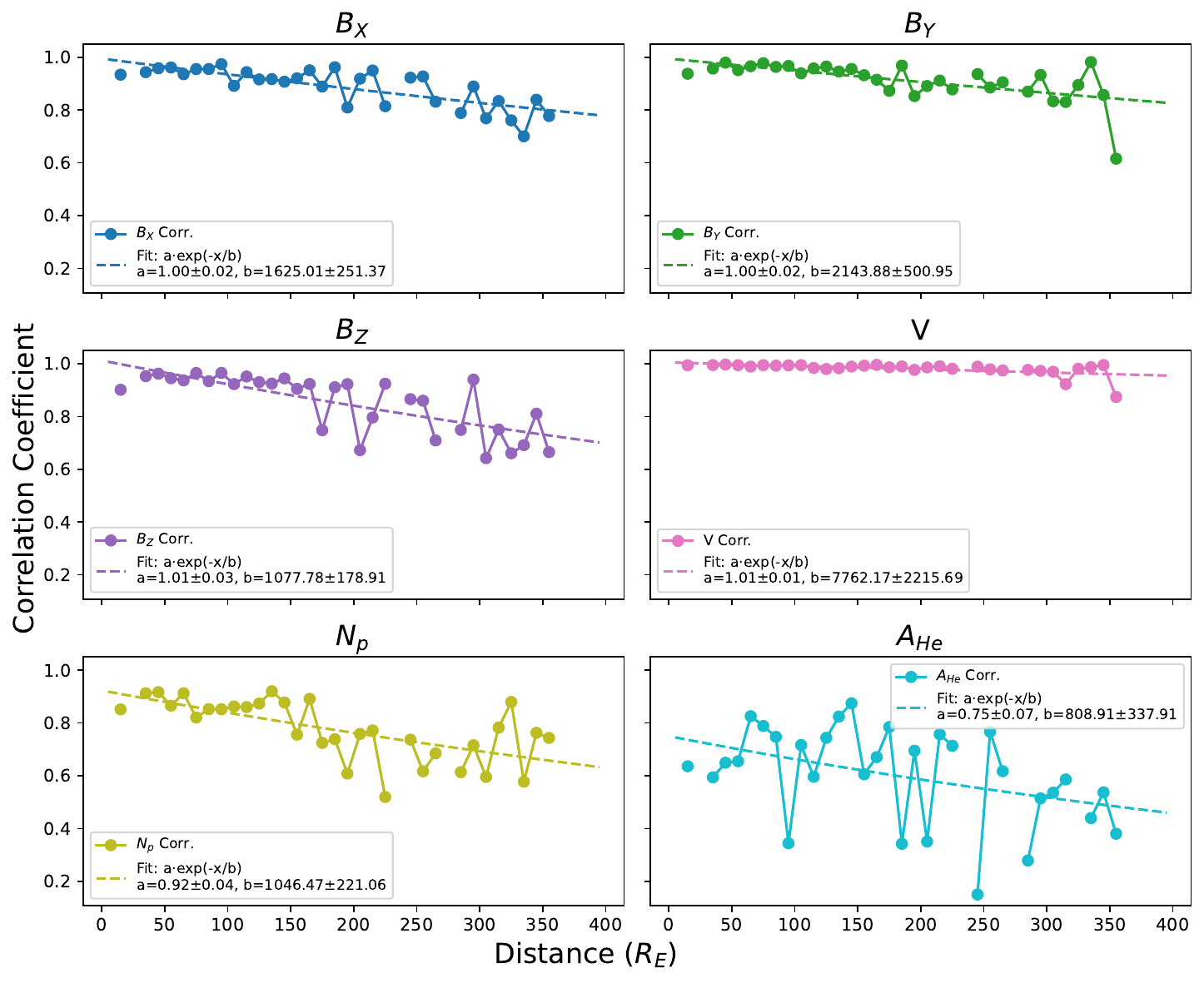}
\caption{Correlation evolution of plasma and magnetic field parameters with distance during ICME intervals. Each panel shows the correlation coefficient as a function of distance (in $R_E$) for different parameters: $B_x$, $B_y$, $B_z$, $V$, $N_P$, and $A_{He}$. The solid line represents an exponential fit of the form $a \cdot \exp(-x/b)$, where $a$ and $b$ are the best-fit parameters indicated in each panel.}
\label{fig:icmes}
\end{figure}

We apply the similar spatial correlation analysis to intervals classified as ICMEs, using simultaneous measurements from the \textit{ACE} and \textit{Wind} spacecraft as a function of their instantaneous separation distance. This is shown in \ref{fig:icmes}.

During ICME intervals, the interplanetary magnetic field (IMF) components exhibit substantially enhanced spatial coherence compared to the ambient solar wind, consistent with the presence of large-scale, organized magnetic structures. All three IMF components display near-unity zero-separation correlations ($a \approx 1$), indicating a high degree of magnetic uniformity at small separations. The transverse component ($B_y$) shows the largest coherence scale, with $b = 2143.88 \pm 500.95~R_E$, followed by the radial component ($B_x$) with $b = 1625.01 \pm 251.37~R_E$. The north--south component ($B_z$) decorrelates more rapidly than $B_x$ and $B_y$, but still maintains a substantial coherence length of $b = 1077.78 \pm 178.91~R_E$, significantly larger than in the background solar wind.

The bulk solar wind speed ($V$) remains highly coherent within ICMEs, with $a = 1.01 \pm 0.01$ and $b = 7762.17 \pm 2215.69~R_E$, reflecting the relatively uniform motion of ICME plasma over extended distances. In contrast, the proton number density ($N_p$) exhibits a reduced zero-separation correlation of $a = 0.92 \pm 0.04$ and a shorter de-correlation length of $b = 1046.47 \pm 221.06~R_E$, indicating increased spatial variability in density within ICME structures. The alpha-to-proton abundance ratio ($A_{\mathrm{He}}$) shows the weakest spatial coherence, with $a = 0.75 \pm 0.07$ and $b = 808.91 \pm 337.91~R_E$, suggesting compositional inhomogeneities and substructure within ICME plasma. The another possible reason behind this can be the instrumental measurement capabilities. 

Overall, these results demonstrate that ICMEs are characterized by substantially larger magnetic-field coherence scales than the background solar wind, consistent with their interpretation as magnetically dominated, coherent flux-rope-like structures.

\subsection{Correlation Between Wind and ACE Observation as a Function of Spacecraft Separation in the SIRs}

\begin{figure}[h!]
\centering
\includegraphics[scale=0.6]{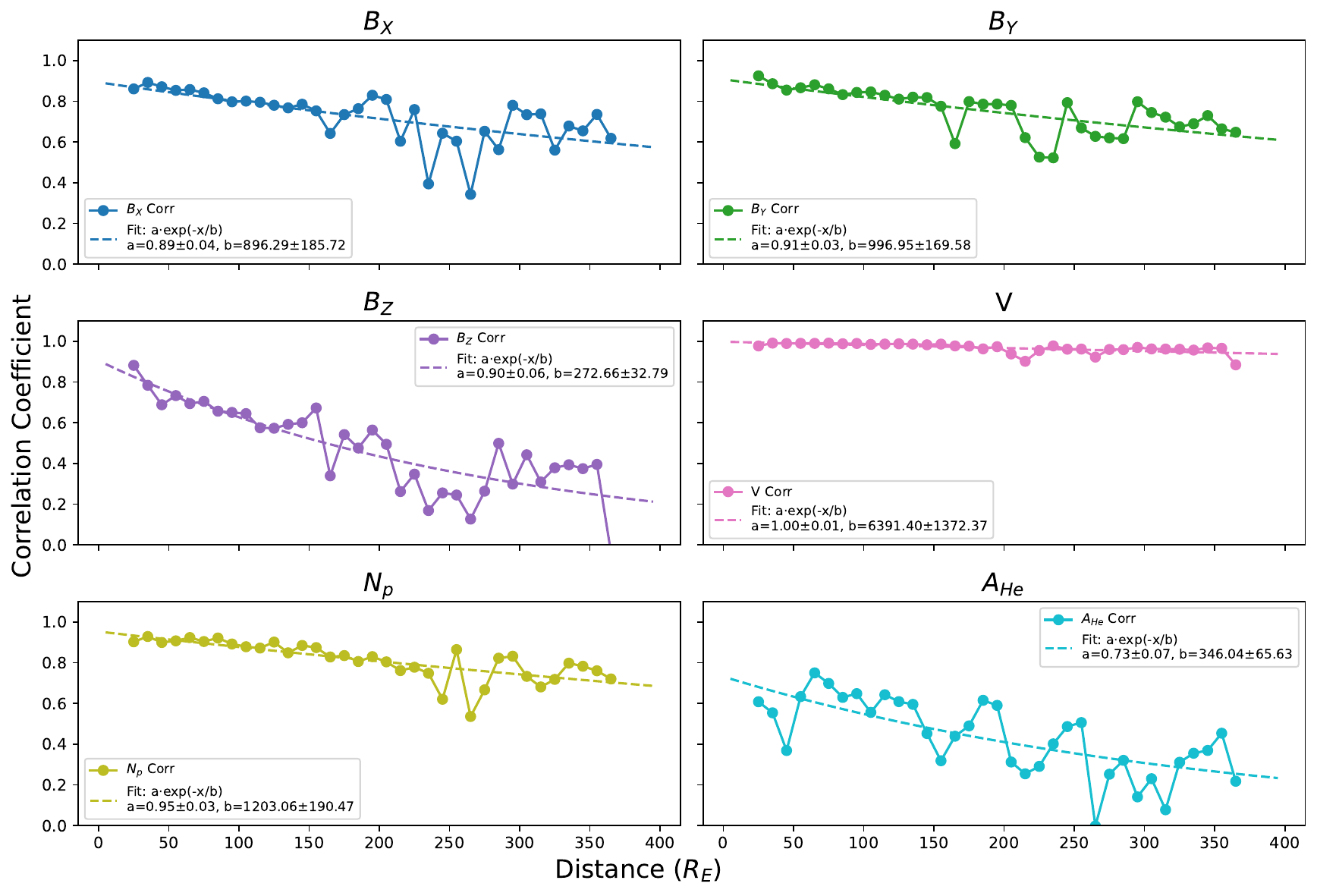}
\caption{Correlation evolution of plasma and magnetic field parameters with distance during SIR intervals. Panels show the variation of correlation coefficient with distance (in $R_E$) for $B_x$, $B_y$, $B_z$, $V$, $N_P$, and $A_{He}$. Each plot includes an exponential fit $a \cdot \exp(-x/b)$ with corresponding fit parameters shown inside the panels.}
\label{fig:sirs}
\end{figure}

Similar to previous two subsection, we next examine the spatial coherence of solar wind parameters during SIR intervals using simultaneous observations from the \textit{ACE} and \textit{Wind} spacecraft. This is shown in \ref{fig:sirs}.

During SIR intervals, the interplanetary magnetic field (IMF) components exhibit moderate spatial coherence. The $B_x$ and $B_y$ components show comparable behavior, with zero-separation correlations of $a = 0.89 \pm 0.04$ and $a = 0.91 \pm 0.03$, and de-correlation lengths of $b = 896.29 \pm 185.72~R_E$ and $b = 996.95 \pm 169.58~R_E$, respectively. In contrast, the $B_z$ decorrelates much more rapidly, with a substantially shorter coherence scale of $b = 272.66 \pm 32.79~R_E$, reflecting enhanced small-scale variability and strong magnetic gradients typically associated with compressed and turbulent SIR plasma.

The bulk solar wind speed ($V$) remains highly coherent during SIRs, exhibiting a zero-separation correlation of $a = 1.00 \pm 0.01$ and a large de-correlation length of $b = 6391.40 \pm 1372.37~R_E$. This reflects the persistent, large-scale velocity structures inherent to SIRs, arising from interactions between fast and slow solar wind streams. The proton number density ($N_p$) also maintains relatively strong spatial coherence, with $a = 0.95 \pm 0.03$ and $b = 1203.06 \pm 190.47~R_E$, consistent with extended compression regions.

In contrast, the alpha-to-proton abundance ratio ($A_{\mathrm{He}}$) exhibits the weakest spatial coherence among the parameters considered, with $a = 0.73 \pm 0.07$ and a short de-correlation length of $b = 346.04 \pm 65.63~R_E$. This indicates that compositional variations within SIRs occur on relatively small spatial scales, likely reflecting local mixing processes and source-region contributions.

Overall, SIRs display spatial coherence properties that are distinct from both the background solar wind and ICMEs. While bulk flow speed and proton density remain coherent over large distances, the IMF—particularly the $B_z$ component—and plasma composition exhibit significantly reduced coherence scales, consistent with the compressed, sheared, and turbulent nature of SIR plasma.

\subsection{Comparison Across Solar Wind Regimes}

Table~\ref{tab:correlation_fits} summarizes the fitted exponential parameters for the spatial correlation of six key solar wind parameters across three regimes: background solar wind, ICMEs, and SIRs. The results reveal systematic differences in spatial coherence that reflect the underlying structure and dynamics of each regime.

The interplanetary magnetic field (IMF) components show the most pronounced variation. ICMEs exhibit the largest coherence scales for all three components, with near-unity zero-separation correlations and de-correlation lengths substantially exceeding those in both the background solar wind and SIRs. This behavior is consistent with the interpretation of ICMEs as large-scale, magnetically organized structures. The contrast is most pronounced for the $B_z$ component, which decorrelates rapidly in the background solar wind ($b \sim 358~R_E$) and SIRs ($b \sim 273~R_E$) but remains coherent over much larger distances within ICMEs ($b \sim 1078~R_E$). SIRs display the shortest coherence lengths for the IMF, particularly in $B_z$, reflecting the compressed, sheared, and turbulent nature of interaction regions. Both $B_x$ and $B_y$ follow similar trends, with ICMEs showing the largest coherence scales, SIRs the smallest, and background solar wind intermediate.

\begin{table*}[h!]
\centering
\caption{Fit parameters ($a$, $b$) from the exponential model $a \cdot \exp(-x/b)$ for correlation coefficients as a function of distance for different solar wind structures: ICME, WIND, and SIR. Values are shown as $a \pm \Delta a$, $b \pm \Delta b$.}
\begin{tabular}{|l|cc|cc|cc|}
\hline
\textbf{Parameters} & \multicolumn{2}{c|}{\textbf{ICME}} & \multicolumn{2}{c|}{\textbf{WIND}} & \multicolumn{2}{c|}{\textbf{SIR}} \\
\cline{2-7}
 & $a$ & $b$ & $a$ & $b$ & $a$ & $b$ \\
\hline
$B_x$  & $1.00 \pm 0.02$ & $1625.01 \pm 251.37$ & $0.91 \pm 0.02$ & $1130.89 \pm 165.84$ & $0.89 \pm 0.04$ & $896.29 \pm 185.72$ \\
\hline
$B_y$  & $1.00 \pm 0.02$ & $2143.88 \pm 500.95$ & $0.92 \pm 0.02$ & $1103.72 \pm 137.23$ & $0.91 \pm 0.03$ & $996.95 \pm 169.58$ \\
\hline
$B_z$  & $1.01 \pm 0.03$ & $1077.78 \pm 178.91$ & $0.87 \pm 0.05$ & $357.83 \pm 43.43$  & $0.90 \pm 0.06$ & $272.66 \pm 32.79$ \\
\hline
$V$    & $1.01 \pm 0.01$ & $7762.17 \pm 2215.69$ & $1.00 \pm 0.00$ & $8673.87 \pm 1143.46$ & $1.00 \pm 0.01$ & $6391.40 \pm 1372.37$ \\
\hline
$N_P$  & $0.92 \pm 0.04$ & $1046.47 \pm 221.06$ & $0.94 \pm 0.02$ & $1711.04 \pm 236.85$ & $0.95 \pm 0.03$ & $1203.06 \pm 190.47$ \\
\hline
$A_{He}$ & $0.75 \pm 0.07$ & $808.91 \pm 337.91$ & $0.83 \pm 0.03$ & $644.67 \pm 93.40$  & $0.73 \pm 0.07$ & $346.04 \pm 65.63$ \\
\hline
\end{tabular}\label{tab:correlation_fits}
\end{table*}

Among the plasma parameters, the bulk solar wind speed ($V$) remains highly coherent across all regimes, with de-correlation lengths of several thousand $R_E$. The largest coherence is observed in the background solar wind ($b \sim 8674~R_E$), slightly reduced in ICMEs ($b \sim 7762~R_E$) and SIRs ($b \sim 6391~R_E$). This indicates that large-scale velocity structures persist over extended distances even within dynamically evolving transients. Proton number density ($N_p$) shows intermediate behavior, remaining well correlated in all regimes but with the longest coherence in the background solar wind ($b \sim 1711~R_E$), reflecting smoother large-scale density variations outside of organized structures, and somewhat shorter coherence in ICMEs ($b \sim 1046~R_E$) and SIRs ($b \sim 1203~R_E$) due to compression and structuring.

The alpha-to-proton abundance ratio ($A_{\mathrm{He}}$) consistently exhibits the weakest spatial coherence. Zero-separation correlations are lower, and de-correlation lengths are shorter in all regimes, particularly in SIRs ($b \sim 346~R_E$), suggesting that compositional variations occur on smaller spatial scales and are influenced by localized mixing processes or source-region inhomogeneities. Another possible reason could be differences in how the instruments measure the helium abundance.

Overall, these results demonstrate that the spatial coherence of solar wind parameters depends strongly on the underlying solar wind structure. ICMEs are dominated by large-scale magnetic coherence, SIRs by enhanced small-scale variability and reduced magnetic coherence, and the background solar wind exhibits intermediate behavior with especially large-scale coherence in bulk plasma properties such as velocity and density. These findings provide quantitative insight into the characteristic spatial scales of solar wind fluctuations in different regimes, with implications for multi-spacecraft observations and space weather modeling.

\section{Discussion}\label{sec:disc}

The spatial correlation analysis of simultaneous \textit{ACE} and \textit{Wind} observations reveals pronounced differences in coherence scales among solar wind plasma and magnetic field parameters. Previous studies, such as \citet{King2005}, have reported correlations between these parameters, but did not investigate how these correlations vary with spacecraft separation. Most prior work has focused primarily on magnetic field parameters \citep[e.g.,][]{Wicks2009,Wicks2010,Ragot2022} and has been limited to the background solar wind. In contrast, our results provide new insight into the spatial organization of the solar wind and the physical processes governing the evolution of both large-scale structures and small-scale fluctuations.

The exceptionally large de-correlation length obtained for the bulk solar wind speed indicates that the flow velocity is dominated by large-scale, long-lived structures that remain coherent over distances of several thousand Earth radii. This behavior is consistent with the interpretation that solar wind velocity primarily reflects the global expansion of coronal source regions and is only weakly affected by local turbulent fluctuations on the spatial scales sampled by ACE and \textit{Wind}. The near-unity zero-separation correlation further suggests that measurement uncertainties and intrinsic variability play a minimal role in degrading velocity coherence at small separations. Similar large correlation scales (around 0.25 AU) have been reported by \citet{Podesta2008} using STEREO A and B observations.

Proton number density ($N_p$) also exhibits substantial spatial coherence, although its de-correlation length is significantly shorter than that of the bulk speed. Density fluctuations are more susceptible to compressive processes, including pressure-balanced structures, stream interactions, and compressive turbulence, which introduce spatial variability on smaller scales. The intermediate de-correlation length for $N_p$ likely reflects the combined influence of large-scale solar wind structuring and mesoscale compressive fluctuations.

In contrast, the interplanetary magnetic field (IMF) components exhibit considerably shorter coherence lengths, together with a clear anisotropy among the field directions. The radial ($B_x$) and transverse ($B_y$) components show comparable de-correlation scales, suggesting that they are controlled by similar large-scale spatial processes, such as Parker-spiral field geometry, field-line meandering, and Alfvénic fluctuations. At 1 AU, the zeroth-order IMF structure is approximately described by a Parker spiral oriented at $\sim45^\circ$ in the ecliptic plane, resulting in comparable mean magnitudes for $B_x$ and $B_y$. Consequently, the observed correlations in these components are strongly influenced by this large-scale background magnetic structure. In contrast, the north--south component ($B_z$) exhibits a substantially shorter coherence scale, implying dominance by more localized and rapidly evolving structures. Because the average IMF in the ecliptic plane has $B_z \approx 0$, this component lacks a strong large-scale background field and is therefore more sensitive to turbulent fluctuations, discontinuities, and smaller-scale dynamics. The outer scale of solar wind turbulence at 1 AU is approximately $10^6$ km \citep{Howes2008}, which is comparable to the observed solar wind correlation length. As a result, $B_z$ decorrelates much more rapidly with increasing spacecraft separation compared to $B_x$ and $B_y$.

The alpha-to-proton abundance ratio ($A_{\mathrm{He}}$) shows the weakest spatial coherence, with both lower zero-separation correlations and shorter de-correlation lengths. While we expect the compositional correlation to be high since abundances are largely established near the Sun. The reduced coherence likely reflects differences in instrument measurement techniques between \textit{ACE} and \textit{Wind}, as well as localized variability during solar wind transport.

A caveat of this study is that the use of a simple exponential model for the correlation may not fully capture the underlying physics of the system. In particular, spatial structures and distinct plasma regimes such as transitions between different solar wind streams can introduce characteristic scales that strongly influence the observed variability. Once these scales are exceeded, the correlations may decrease more rapidly than predicted by a single-scale exponential decay. Since the present analysis is limited to separations up to $\sim 300,R_E$, while some inferred correlation lengths are larger than this range, the results likely represent only the initial stage of decorrelation rather than the full decay to statistical independence. This limitation is primarily due to observational constraints. A more complete characterization of the correlation behavior would require multi-spacecraft measurements with a wider range of separations and geometries, such as those provided by STEREO missions, and is therefore left for future work.

Taken together, these results demonstrate a clear hierarchy of spatial coherence in the solar wind. Bulk plasma properties, such as flow speed and density, remain coherent over large distances, reflecting their origin in large-scale coronal and heliospheric structures. Magnetic field fluctuations, particularly in the north--south direction, and compositional parameters decorrelate on much shorter scales, consistent with the influence of turbulence, discontinuities, and kinetic processes. The rapid decorrelation of $B_z$ has important implications for space weather applications, as it limits the predictability of geoeffective magnetic field conditions based on upstream measurements from widely separated spacecraft. 

Given that the OMNI dataset relies heavily on \textit{ACE} and \textit{Wind} observations, our results indicate that careful attention must be paid to the spacecraft separation when comparing or using these data for space weather prediction. These findings underscore the value of multi-point observations in characterizing the spatial variability of solar wind parameters and suggest that caution is warranted when extrapolating magnetic field measurements especially $B_z$ across large distances. These results can also aid in the cross‑calibration of plasma instruments on new spacecraft, such as the Aditya Solar Wind Particle Experiment (ASPEX) payload \citep{Goyal2018, Kumar2025, Akumar2025, Sebastian2026} and MAG \citep{Yadav2025} onboard the Aditya‑L1 mission \citep{Seetha2017, Tripathi2022, Parate2025}, which measures solar wind protons and alpha particles and provides high‑quality compositional and plasma data at L1.

\section{Conclusions}\label{sec:conclusion}

In this study, we quantified the spatial coherence of key solar wind plasma and magnetic field parameters using simultaneous \textit{ACE} and \textit{Wind} observations as a function of spacecraft separation, systematically comparing background solar wind, SIRs, and ICMEs. By modeling the decay of Pearson correlation coefficients with an exponential function, we demonstrated that solar wind structures exhibit markedly different coherence scales depending on their underlying physical characteristics. 
Bulk flow speed consistently remains the most spatially coherent parameter across all regimes, whereas plasma composition exhibits the weakest coherence. The interplanetary magnetic field shows strong regime dependence: ICMEs display near-unity correlations and the largest magnetic coherence scales, consistent with large-scale, organized flux-rope-like structures, while SIRs exhibit reduced coherence (particularly in the $B_z$ component) reflecting compressed, sheared, and turbulent plasma. The rapid decorrelation of $B_z$ has important implications for space weather applications, as it limits the predictability of geoeffective magnetic field conditions based on upstream measurements from widely separated spacecraft.  The background solar wind generally exhibits intermediate behavior, with relatively large coherence in bulk plasma properties but shorter coherence lengths in magnetic fluctuations.
These results provide a quantitative framework for distinguishing solar wind structures based on their spatial coherence properties and offer important constraints for multi-spacecraft analyses, and space weather modeling. They also highlight the importance of considering spacecraft separation when comparing measurements and underscore the value of multi-point observations for capturing the full range of spatial variability in the solar wind.

\section*{Acknowledgments}
We gratefully acknowledge the principal investigators of instruments on-board NASA's Wind and ACE mission for generating the data and making them publicly available. Y. acknowledge the support by the College of Liberal Arts and Sciences at the University of Iowa. G.G.H. acknowledges support of NASA grants 80NSSC24K124 and 80NSSC24K0552. We express our sincere gratitude to the Department of Space, Government of India for supporting this work. We acknowledge use of NASA/GSFC’s Space Physics Data Facility (SPDF).

\bibliographystyle{jasr-model5-names}
\biboptions{authoryear}
\bibliography{manuscript}

\end{document}